\documentclass[10pt]{iopart}
\usepackage{graphicx}
\usepackage[latin1]{inputenc}
\usepackage{amssymb}
\usepackage{xcolor}
\usepackage{iopams}

\begin{document}




\title{Geometry induced domain-walls of dipole lattices on curved structures}
\date{\today}

\author{Ansgar Siemens$^1$, Peter Schmelcher$^1$$^2$}

\address{$^1$ Zentrum f\"ur Optische Quantentechnologien, Fachbereich Physik, Universit\"at Hamburg, Luruper Chaussee 149, 22761 Hamburg Germany}
\address{$^2$ Hamburg Center for Ultrafast Imaging, Universit\"at Hamburg, Luruper Chaussee 149, 22761 Hamburg Germany}
\ead{asiemens@physnet.uni-hamburg.de}
\ead{pschmelc@physnet.uni-hamburg.de}

\begin{abstract}
\noindent We investigate the ground state properties of rectangular dipole lattices on curved surfaces. 
The curved geometry can `distort' the lattice and lead to dipole equilibrium configurations that strongly depend on the local geometry of the surface. 
We find that the system's ground state can exhibit domain-walls separating domains with different dipole configurations.  
Furthermore, we show how, regardless of the surface geometry, the domain-walls locate along the lattice sites for which the (Euclidean) distances to nearest and next-nearest neighbors are equal. 
We analyze the response of the domain-walls to an external electric field and observe displacements and splittings thereof below and above a critical electric field, respectively. 
We further show that the domain-wall acts as a boundary that traps low-energy excitations within a domain. 

\end{abstract}

\maketitle
\ioptwocol 

\section{Introduction}

Topology based concepts are widely used for the description of physical systems. 
A prominent example is that of domains and domain-walls (DW) arising from spontaneous symmetry breaking, with applications ranging from optics, where DWs separate differently polarized regions \cite{pitois1998,zhang2009}, to magnetic Bose-Einstein condensates \cite{yu2021}, or even string theory \cite{choi1985}. 
Moreover, in materials or model systems described by lattices of interacting electric dipoles, the ground-state (GS) degeneracy of the dipoles, i.e. the invariance of the energy under inversion of all dipoles, can lead to the formation of local domains with different dipole orientations \cite{nataf2020} separated by DWs. 
For lattices of electric dipoles (related to ferroelectrics (FE)), experiments have shown a great controllability of these domains and DWs, allowing for their artificial creation, annihilation, or controlled shifts \cite{meier2022,sharma2017}. 
This, together with the fact that long-range order of electric dipoles has been found at room temperature in a multitude of materials \cite{yuan2019}, has made FE materials interesting candidates for applications, such as smart sensors, capacitors, transducers, actuators, energy harvesting devices, and non-volatile memories \cite{meier2022,sharma2017,wang2022,muralt2000}. 

It has been shown that the application of strain to a FE material can be used to control the overall FE response, as well as the DW energy and mobility \cite{beckman2009}. 
In some FE materials, the GS dipole configuration can even significantly change when the stress on the material exceeds a (material dependent) critical value - an effect known as ferroelasticity \cite{salje2012,belletti2014,li2022,nataf2020}. 
The sudden change of the GS in ferroelastic materials usually emerges due to (compression-induced) changes of the underlying crystal structure. 
Consequently, the properties of strain-induced DWs in ferroelastic materials can drastically differ from the properties of DWs in FE materials \cite{belletti2014}. 

Similar phenomenology can also be found in magnetic materials \cite{parkin2008}. 
A difference between ferroelectrics and magnets is that the formation of domains in ferroelectrics is usually due to the dipole-dipole interaction, whereas in magnets other interactions, such as exchange processes \cite{white2007}, dominate the equilibrium dipole alignment. 
In this context, one notable interaction arising in some magnetic systems is the Dzyaloshinskii-Moriya (DM) interaction \cite{dzyaloshinsky1958,moriya1960,camley2023}:
it allows for the formation of non-collinear spin structures, such as magnetic skyrmions \cite{rossler2006,tokura2021,bogdanov2020}, which have been proposed for applications \cite{lai2017,luo2018,chauwin2019} e.g. in high density storage devices \cite{kiselev2011,fert2013}.
Studies of the influence of spatial curvature on magnetic materials (e.g. two-dimensional magnetic films on a spherical surface) have shown that curvature can effectively induce a DM type interaction and thereby significantly change the GS dipole configuration \cite{gaididei2014,streubel2016b,volkov2019}. 
Naturally, the question arises: what is the influence of spatial curvature in ferroelectric materials where the material properties are dominated by dipole-dipole interactions.

The first steps in this direction were already done in studies of dipoles confined to cylindrical or helical geometries. 
These studies range from experimental investigations (e.g. of the dipole equilibrium configurations in stacks of BTA (trialkylbenzene-1,3,5-tricarboxamide) molecules \cite{cornelissen2019,urbanaviciute2019}) to model systems, such as Hubbard models with long-range hopping terms \cite{xiong1992,wang1991,stockhofe2015a,stockhofe2016,guo2020}. 
Already in such (comparatively) simple curved geometries, the GS properties show a strong dependence on the geometrical parameters: 
for example, for dipoles with fixed positions along a helical path, it was demonstrated that the GS configurations are described by a complex self-similar bifurcation diagram \cite{siemens2022}. 
Even in systems without anisotropic interactions, such as Coulomb-interacting ions on a helix, the confinement to a curved path or surface embedded in euclidean space induces a plethora of phenomena \cite{schmelcher2011,zampetaki2013,pedersen2014,zampetaki2015,zampetaki2015a,zampetaki2017,zampetaki2018,siemens2020,siemens2021} that are absent in structures without curvature. 

Motivated by the above facts, we focus on possible novel features introduced into model systems consisting of dipole lattices due to their curved structure. 
We start by explaining how the structural arrangement of dipoles influences the GS and can lead to DWs that are `pinned' through geometrical parameters of the system. 
We investigate the static DW properties and their dependencies on system parameters for a two-dimensional prototype system, and highlight the differences to those DWs that are commonly observed in ferroelectric materials by showing the reaction of our DW to applied external fields. 
Then, using the example of a toroidal dipole lattice, we demonstrate and discuss the appearance of the previously introduced properties. 
It is further shown, how these DWs act as barriers that prevent low energy excitations from traveling freely through the system. 

This work is structured as follows: 
Our setup and methodology for the dipoles on curved or deformed surfaces is described in Sec. \ref{Sec: General Setup}. 
Section \ref{Sec:2D_lattice} then explains the physics of domains and DWs on curved dipole lattices using a simple two-dimensional example system. 
In Sec. \ref{Sec:TorDipLatt}, we explore and analyze the toroidal dipole lattice. 
Sec. \ref{Sec:Excitations} investigates the eigenmodes based on the GS of the toroidal dipole lattice and we show that DWs can act as `barriers' preventing excitations from traveling through the system. 
Finally, our results are summarized in Sec. \ref{Sec:Summary}.

\section{General Setup and Methodology}
\label{Sec: General Setup}

\begin{figure}
\includegraphics[width=\columnwidth]{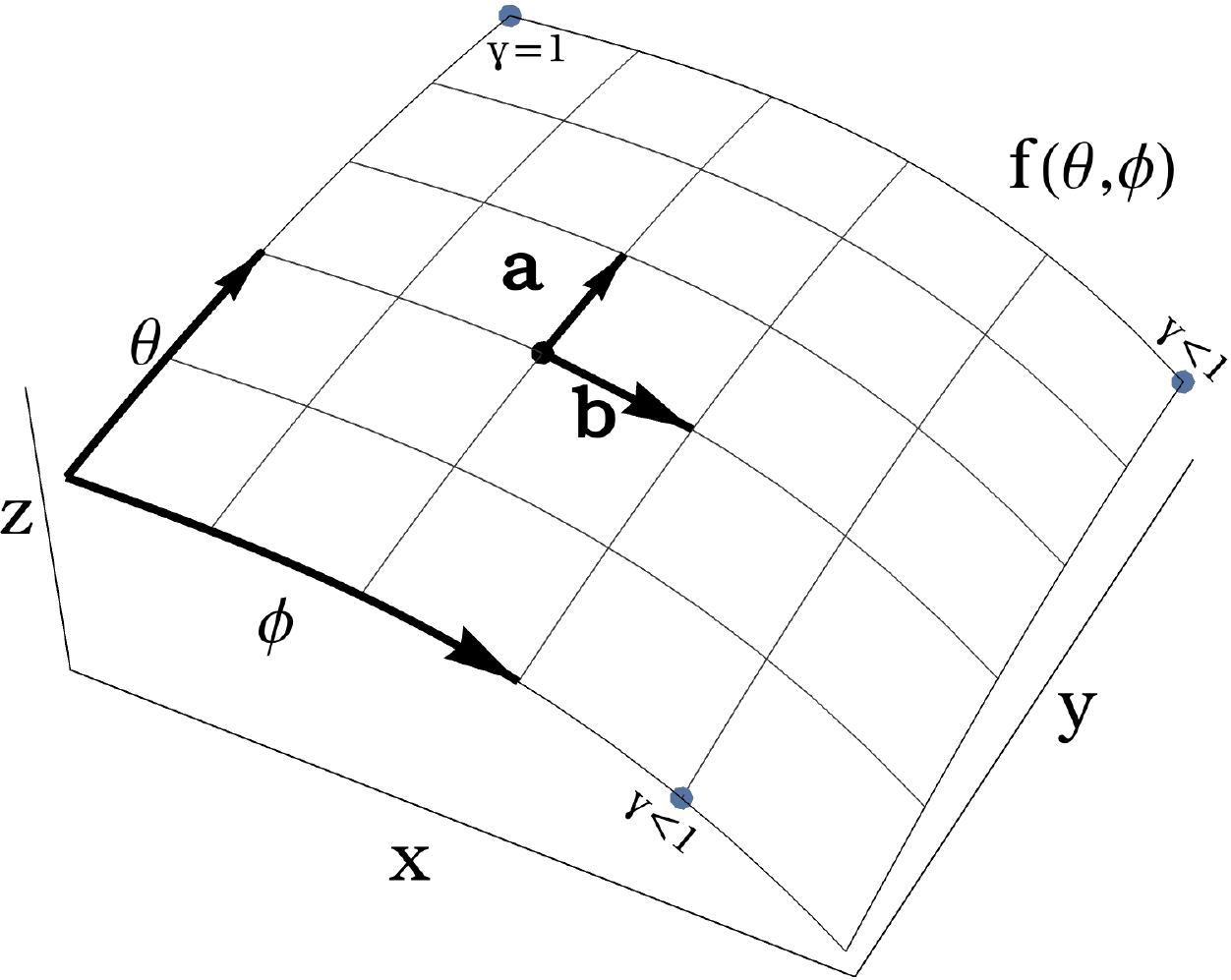}
\caption{\label{Fig:Setup} Example of a lattice spanned on a curved surface $f(\theta,\phi)$. As a result of the curvature, the euclidean distances $\textbf{a}(\theta,\phi)$ and $\textbf{b}(\theta,\phi)$ between lattice points depend on the local geometry of the surface. 
}
\end{figure}

We investigate the physics of dipoles placed on rectangular Bravais lattices spanned on the surface of curved structures (compare Fig. \ref{Fig:Setup}). 
In general, a curved surface, i.e., each point on it, can be completely described by a parametric function $f(\theta,\phi)$ depending on two parametric coordinates $\theta$ and $\phi$ (being internal coordinates of the surface, not necessarily angles). 
For convenience we introduce $\textbf{\^{e}}_{\theta}$ and $\textbf{\^{e}}_{\phi}$ as unit vectors that span a basis in the parametric coordinates. 
These basis vectors are used to indicate directions along the surface. 
Without loss of generality, we define our lattice points to be equidistant in the parametric coordinates, with the parametric distances $\Delta\theta$ and $\Delta\phi$. 
This way, the euclidean distances, namely $a=|\textbf{a}|$ and $b=|\textbf{b}|$ (see Fig. \ref{Fig:Setup}), between neighboring lattice points along $\textbf{\^{e}}_{\theta}$ and $\textbf{\^{e}}_{\phi}$ become dependent on the local geometry of the surface,
\begin{equation}
\begin{array}{c}
\end{array}
\eqalign{
a(\theta,\phi):=& || f(\theta+\Delta\theta,\phi)-f(\theta,\phi) || \cr
b(\theta,\phi):=& || f(\theta,\phi+\Delta\phi)-f(\theta,\phi) || \ .
}
\end{equation}

This dependence of $a$ and $b$ on the local surface geometry will have immediate consequences on the arrangement (such as the orientation) of the dipoles placed in such a lattice. 
For instance at one point on the surface the (euclidean) nearest neighbor (NN) of a dipole is reached through translation along $\textbf{\^{e}}_{\theta}$, while at another point the NN may be found along $\textbf{\^{e}}_{\phi}$. 
An example for this is shown in Fig. \ref{Fig:Setup}, where the ratio $\gamma=a/b$ is one for the lattice point in the uppermost part of the figure (indicated by the blue dot marked $\gamma=1$), but visibly reaches values $\gamma < 1$ for the lattice points in the right and lower part of the figure (indicated by the blue dots marked $\gamma<1$). 
As we will show in Sec. \ref{Sec:2D_lattice}, the properties of the many-body ground state (GS) of dipoles in such a curved lattice depends mainly on this spatially varying ratio $\gamma$. 

The dipoles on the lattice points interact with dipole-dipole interactions, and are in a second step exposed to an external electric field $E$. 
With this, the potential energy of the $n$-th dipole in a lattice containing $K$ dipoles is given by 
\begin{equation}\label{eq:potentialEnergy}
V_n = \sum_{i=1, i\neq n}^{K} \frac{1}{4\pi}\left[\frac{\textbf{d}_i\,\textbf{d}_n}{r_{in}^3}-\frac{3(\textbf{d}_i\cdot\textbf{r}_{in})(\textbf{d}_n\cdot\textbf{r}_{in})}{r_{in}^5}\right]
+\textbf{d}_n\cdot\textbf{E}
\end{equation}
where $\textbf{d}_i$ and $\textbf{d}_n$ are the dipole moments of the dipoles at lattice sites $i$ and $n$ respectively, and $\textbf{r}_{in}$ denotes the (euclidean) vector between them. 

For our computational approach to determine the GS configurations, we use local polar coordinates to describe the orientation of each dipole. 
Furthermore, since a change in the magnitude $d=|\textbf{d}|$ of the dipoles only leads to a scaling of the interaction term in Eq. \ref{eq:potentialEnergy}, we will without loss of generality use $d=1$.
The dipole equilibrium configurations are determined using a principle axis method \cite{brent2002algorithms} - a derivative-free optimization method that performs line-search optimizations along a set of (continuously updated) conjugate search directions. 
Specifically, we use Brent's algorithm which ensures linear independence of the search directions after they are updated. 
Unless explicitly stated otherwise, the results shown in the following correspond to GS configurations. 
Additionally, we use a NN approximation, i.e., considering only interactions with the $4$ nearest dipoles. 
This approximation is valid as long as the lattice is dense enough to locally be considered flat. 
No qualitative differences were observed when comparing the NN approximations to all-to-all simulations. 
The validity of these approximations and the impact of interactions with more distant neighbors are discussed at appropriate places in the following.

\section{Domain-walls in deformed lattices}
\label{Sec:2D_lattice}

In the limit of a vanishing curvature (i.e., a flat surface) the GS dipole configuration is well known \cite{brankov1987,feldmann2008}: 
The dipoles will align along head-to-tail chains with their NNs. 
In a rectangular (non-square) lattice, dipoles in neighboring chains (chains being defined along the shorter NN distance) will orient anti-parallel to one another in an overall anti-ferroelectric (AFE) configuration. 
Thus, the GS of a flat rectangular lattice cannot be a ferroelectric (FE) configuration \cite{brankov1987,rozenbaum1991} when only dipole-dipole interactions are considered. 
In the special case of a square lattice ($\gamma=1$) the GS becomes highly degenerate \cite{debell1997}. 
Flat regions in the potential landscape allow for continuous transformations between these different GS configurations.
The behavior is, however, entirely different for dipole arrays spanned on the surface of a curved or deformed structure. 
This section is dedicated to showing and discussing the behavior emerging when the lattice geometry becomes deformed. 
An application to a specific case, namely the toroidal dipole lattice is provided in Sec. \ref{Sec:TorDipLatt}.

\begin{figure}
\includegraphics[width=\columnwidth]{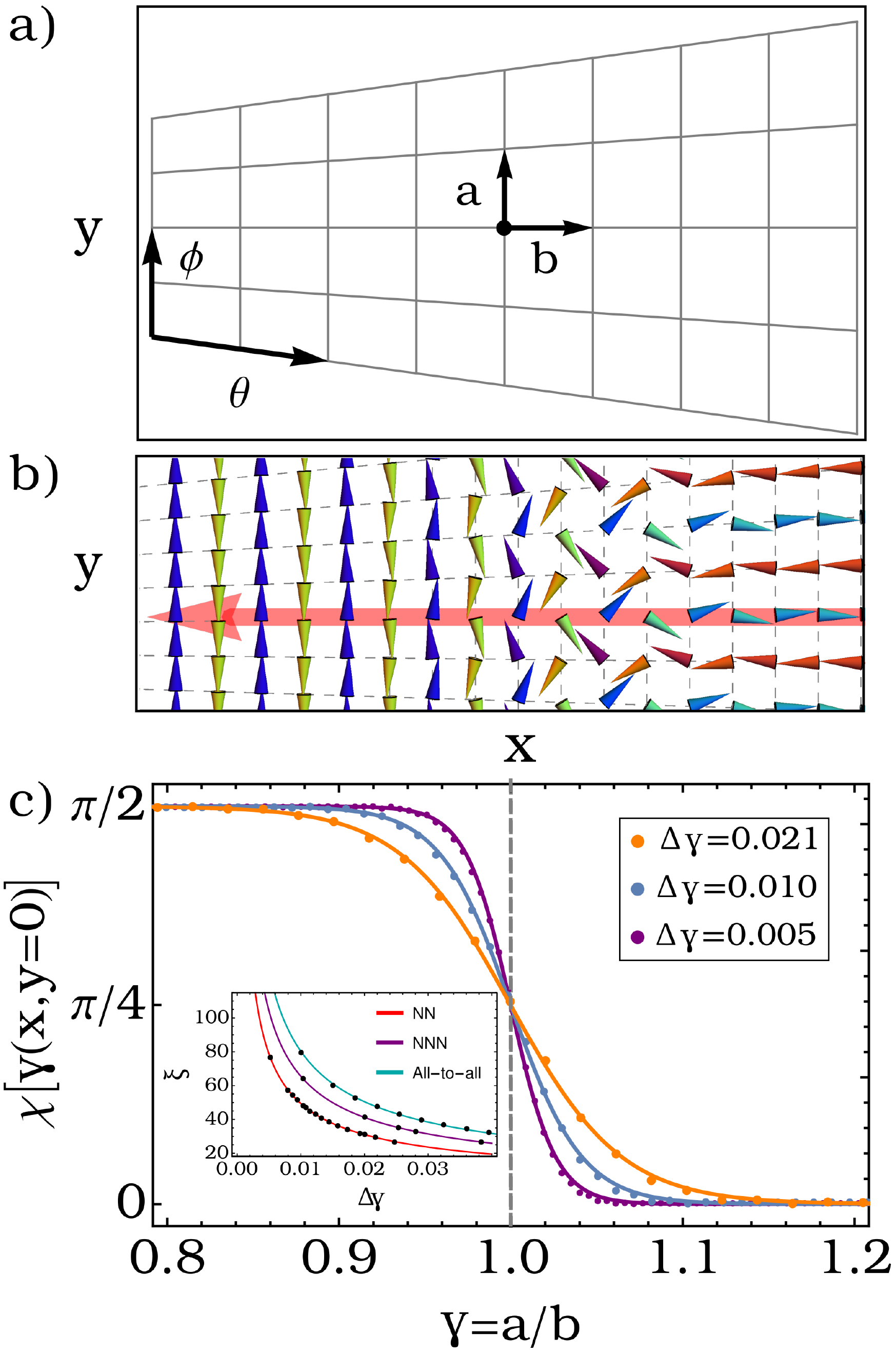}
\caption{\label{Fig:2dSetup} (a) 
Schematic visualization of a lattice spanned on the surface defined by Eq. \ref{eq:2D_surface} together with the lattice defined by Eq. \ref{eq:2DLatticePositions}. 
(b) GS configuration of dipoles on the lattice defined by Eq. \ref{eq:2DLatticePositions} showing two domains separated by a DW. The red arrow indicates a one-dimensional cut through the lattice running perpendicular to the DW. (c) DW profiles obtained from the dipoles located along the one-dimensional cut (red arrow in (b)). $\chi$ describes the angle between the dipoles and the direction of the one-dimensional cut. $\gamma=a/b$ is the corresponding (local) ratio of the euclidean distances to the NNs. The orange, blue, and purple curves are obtained for lattices with $\Delta\gamma=0.021$, $0.01$, and $0.005$ respectively - with $\Delta\gamma$ being defined as the change of $\gamma$ between two successive dipoles along the one-dimensional cut. The inset depicts the dependency of the sigmoid steepness $\xi$ on $\Delta\gamma$ for calculations with nearest (red), next-nearest neighbors (purple), and all-to-all interactions (green). For a detailed discussion, see main text.
}
\end{figure}

A simple flat geometry showcasing the impact of deformation is given by the following parametric surface 
\begin{equation}\label{eq:2D_surface}
f_{2D}(\theta,\phi):= 
\left(
\begin{array}{c}
\theta \\
\theta\cdot\tan[\phi] \\
0
\end{array}
\right)
\end{equation}
where $-\pi/2<\phi<\pi/2$. 
Adopting the nomenclature introduced in Sec. \ref{Sec: General Setup}, we place dipoles on the surface while maintaining constant distances $\Delta\theta$ and $\Delta\phi$ in the parametric coordinates. 
Note that $\Delta\phi$ is considered to be small, i.e., $\Delta\phi\ll\pi/2$, in order to provide a sufficiently large number of lattice points along $\textbf{\^{e}}_{\phi}$. 
The system now describes a `lattice' which narrows with decreasing $\theta$. 
The positions of the lattice points for dipoles on a $(N\times (2M+1))$ grid are given by
\begin{equation}\label{eq:2DLatticePositions}
f_{2D}^{ik}:=
\left(
\begin{array}{c}
i\Delta\theta \\
i\Delta\theta\cdot\tan\left[k\Delta\phi\right] \\
0
\end{array}
\right)
\mbox{  for  }
\left\{
\eqalign{i &\in [1,N] \cr k &\in [-M,M]}
\right.
\end{equation}
We also restrict the dipoles to rotate in the plane spanned by Eq. \ref{eq:2D_surface}. 
A schematic visualization of the lattice based on Eq. \ref{eq:2DLatticePositions} is shown in Fig. \ref{Fig:2dSetup}(a). 
Note that the surface defined in Eq. \ref{eq:2D_surface} possesses a singularity and will map all lattice points for $\theta=0$ (independent of $\phi$) to the same position in euclidean space. 
We will therefore only consider lattice points in the regime $\theta>0$.

As mentioned in Sec. \ref{Sec: General Setup}, the GS dipole orientation mainly depends on $\gamma=a/b$. 
An interesting observation is that, for the lattice spanned by Eq. \ref{eq:2DLatticePositions}, $\gamma$ becomes independent of $\Delta\theta$. 
This is because both $\textbf{a}$ and $\textbf{b}$ scale linearly with $\Delta\theta$. 
It can be directly shown that $\gamma\rightarrow0$ when $\theta\rightarrow\infty$. 
Furthermore, it can be shown that for any given point $f_{2D}^{ik}$ in the lattice one can get $\gamma>1$ when $\Delta\phi$ is chosen sufficiently small. 
We know that in regions where $\gamma>1$ the most favorable alignment of the dipoles will be head-to-tail chains along $\textbf{\^{e}}_{\phi}$. 
On the contrary, if $\gamma<1$ the chains will preferably align along $\textbf{\^{e}}_{\theta}$.  
This implies - provided $\Delta\phi$ is sufficiently small - the presence of two domains with different dipole alignments in the GS. 
This results in the existence of a DW acting as a boundary between the two domains. 
Such a ground state configuration exhibiting the two domains, as well as the separating DW, is shown in Fig. \ref{Fig:2dSetup}(b). 
As expected, the two domains for small and large values of $\theta$ feature AFE dipole configurations aligned along $\textbf{\^{e}}_{\phi}$ and $\textbf{\^{e}}_{\theta}$ respectively. 
To characterize the DW, we focus on those dipoles located along a one-dimensional (1D) cut through the lattice. 
The cut is taken at constant $\phi=0$, perpendicular to the DW. 
In Fig. \ref{Fig:2dSetup}(b) this cut is indicated by the red arrow. 
As an order parameter, we choose the (absolute) angle $\chi$ between the dipoles and the direction of the 1D cut. 
\begin{equation}
\chi=|\arccos\left(\textbf{d}\cdot\textbf{\^{e}}_{\theta}|_{\phi=0}\right)|,
\end{equation}
where $\textbf{\^{e}}_{\theta}|_{\phi=0}$ is the unit vector along $\textbf{\^{e}}_{\theta}$ for $\phi=0$. 
In the domain for large $\theta$ [right of the DW in Fig. \ref{Fig:2dSetup}(b)] we approach $\chi=0$, whereas in the domain for small $\theta$ values we approach $\chi=\pi/2$. 
The angle $\chi$ as a function of the local value of $\gamma$ for all dipoles along the 1D cut is shown in Fig. \ref{Fig:2dSetup}(c) (orange-colored points). 
The blue and purple colored points correspond to calculations with lattices where the change of $\gamma$ along the 1D cut (denoted by $\Delta\gamma$) is smaller than in the orange curve. 
The solid lines correspond to a fitting sigmoid function given by
\begin{equation}\label{Eq:sigmoid}
\Phi(\gamma) = \frac{\pi}{2+2 e^{-\xi(\gamma(\theta)-\gamma_0)}},
\end{equation}
where $\xi$ is the sigmoid steepness and $\gamma_0$ is the value of $\gamma$ at the midpoint of the sigmoid function. 
From now on, we will refer to this midpoint as the position of the DW center. 
In the DW profiles shown in Fig. \ref{Fig:2dSetup}(c), the DW center is located (almost exactly) at $\gamma=1$. 
The deviation from $\gamma=1$ is of the order of $\sim0.01\Delta\theta$ - only a fraction of the lattice constant $\Delta\theta$. 
We determined DW profiles for a wide range of $\Delta\phi$ and did not observe any significant deviation of the DW center from $\gamma=1$. 
Note, that the exact location of the DW at $\gamma=1$ was only observed for NN interactions. 
When all-to-all interactions are considered, a shift of the DW center towards $\gamma<1$ was observed. 
The largest observed shift of the DW center from $\gamma=1$ was on the order of $\sim\Delta\theta$. 
From Fig. \ref{Fig:2dSetup}(c) one can see that the DW steepness (corresponding to the fit-parameter $\xi$ in Eq. \ref{Eq:sigmoid}) is larger for those lattices where $\Delta\gamma$ is smaller: 
A comparison of the fit parameter $\xi$ for different values of $\Delta\gamma$ is shown in the inset of Fig. \ref{Fig:2dSetup}(c) for NN, next-NN (NNN), and all-to-all simulations. 
The only significant difference between the NN, NNN, and all-to-all data is that the DWs are consistently narrower when more interactions are considered. 
Consequently, DWs obtained for a given $\Delta\gamma$ from NN and NNN calculations are wider than corresponding DWs calculated for the same $\Delta\gamma$ with all-to-all interactions. 
Besides this, no qualitative differences could be observed for the DWs.  
For NN, NNN and all-to-all interactions the DW steepness $\xi$ scales (approximately) with $\xi\sim(\Delta\gamma)^{-0.67}$ [purple, red, and green fit-functions in the inset of Fig. \ref{Fig:2dSetup}(c)].

\begin{figure}
\includegraphics[width=\columnwidth]{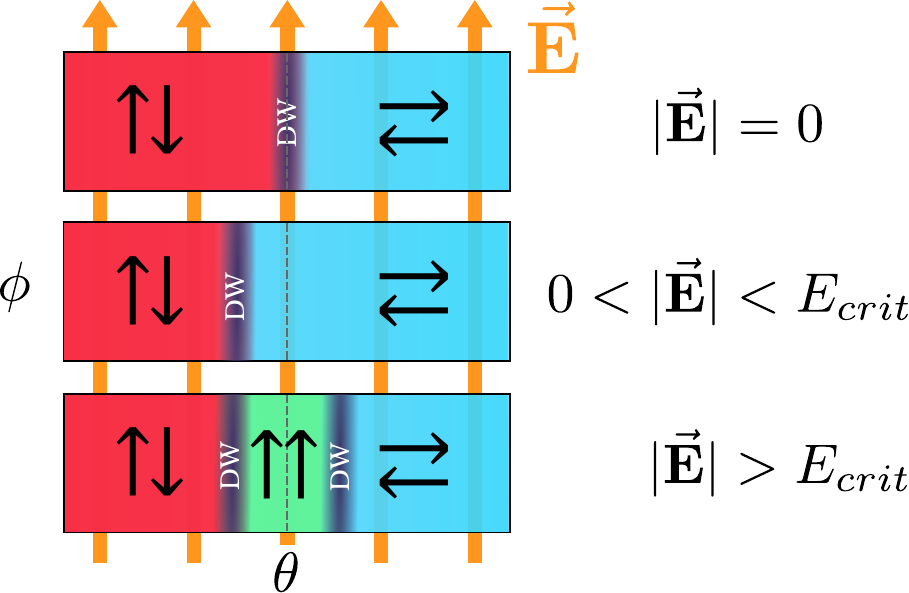}
\caption{\label{Fig:2dField} 
Schematic explaining how the equilibrium dipole orientations change around $\gamma=1$ when an electric field is applied along the y-direction (parallel to $\textbf{\^{e}}_{\phi}$). For $|\vec{E}|=0$, there are two domains with dipoles aligned along $\textbf{\^{e}}_{\theta}$ and $\textbf{\^{e}}_{\phi}$ (respectively corresponding to the blue and red regions) separated by a DW (purple area between the domains). When $|\vec{E}|$ is increased, the domain with dipoles aligned along $\textbf{\^{e}}_{\theta}$ expands, while the domain with dipoles aligned along $\textbf{\^{e}}_{\phi}$ shrinks. As a result, the DW is shifted. When the field is increased further - above a critical value $E_{crit}$ - a new ferroelectric domain with dipoles aligned parallel to the field (green area in the lowest panel) emerges. 
}
\end{figure}

The above discussed geometry induced DW differs significantly from other known types of ferroelectric DWs: 
Firstly, it is a feature of the systems GS (i.e., not an excitation) with a (fixed) position that is predetermined by the geometry of the underlying lattice and surface. 
A second difference concerns the overall response to applied external electric fields. 
To illuminate the latter, we briefly address the impact of a field applied along the y-direction (parallel to $\textbf{\^{e}}_{\phi}$) using an adiabatic method: 
the field strength is increased stepwise, and at each step the configuration is relaxed with a Newton method that takes into account NNN interactions.

A schematic demonstrating the effects of the external field on the dipole orientations in the vicinity of the DW (i.e., in the vicinity of $\gamma=1$) is shown in Fig. \ref{Fig:2dField}. 
The behavior can be summarized as follows: 
For weak fields, the DW separating the two domains shifts to a new equilibrium position. 
For strong fields, a (new) third domain emerges (green area in Fig. \ref{Fig:2dField}). 
The dipoles in this third domain are aligned parallel to the field. 
This third domain will expand with increasing field strength towards both lower and larger values of $\theta$. 
We can explain this behavior in the context of dipoles in square (non-deformed) Bravais lattices (with a global $\gamma=1$): 
In such square lattices, the external field breaks the GS degeneracy and orients the dipoles along the field lines - in an AFE (FE) configuration for weak (strong) fields \cite{klymenko1990}. 
In other words: in a lattice with $\gamma=1$ the dipoles do not have a preferred alignment direction (w.r.t. the primitive lattice vectors) and therefore require on average (compared to lattices with $\gamma\neq1$) weaker fields to align all dipoles in a FE configuration along an arbitrary direction. 
This is consistent with the behavior observed in Fig. \ref{Fig:2dField}. 
Namely, for strong fields the dipoles around $\gamma=1$ form a FE domain parallel to the field - even though the field is not strong enough to significantly affect the dipole orientations in the neighboring domains.

\section{Toroidal dipole lattice}\label{Sec:TorDipLatt}

\begin{figure}
\includegraphics[width=\columnwidth]{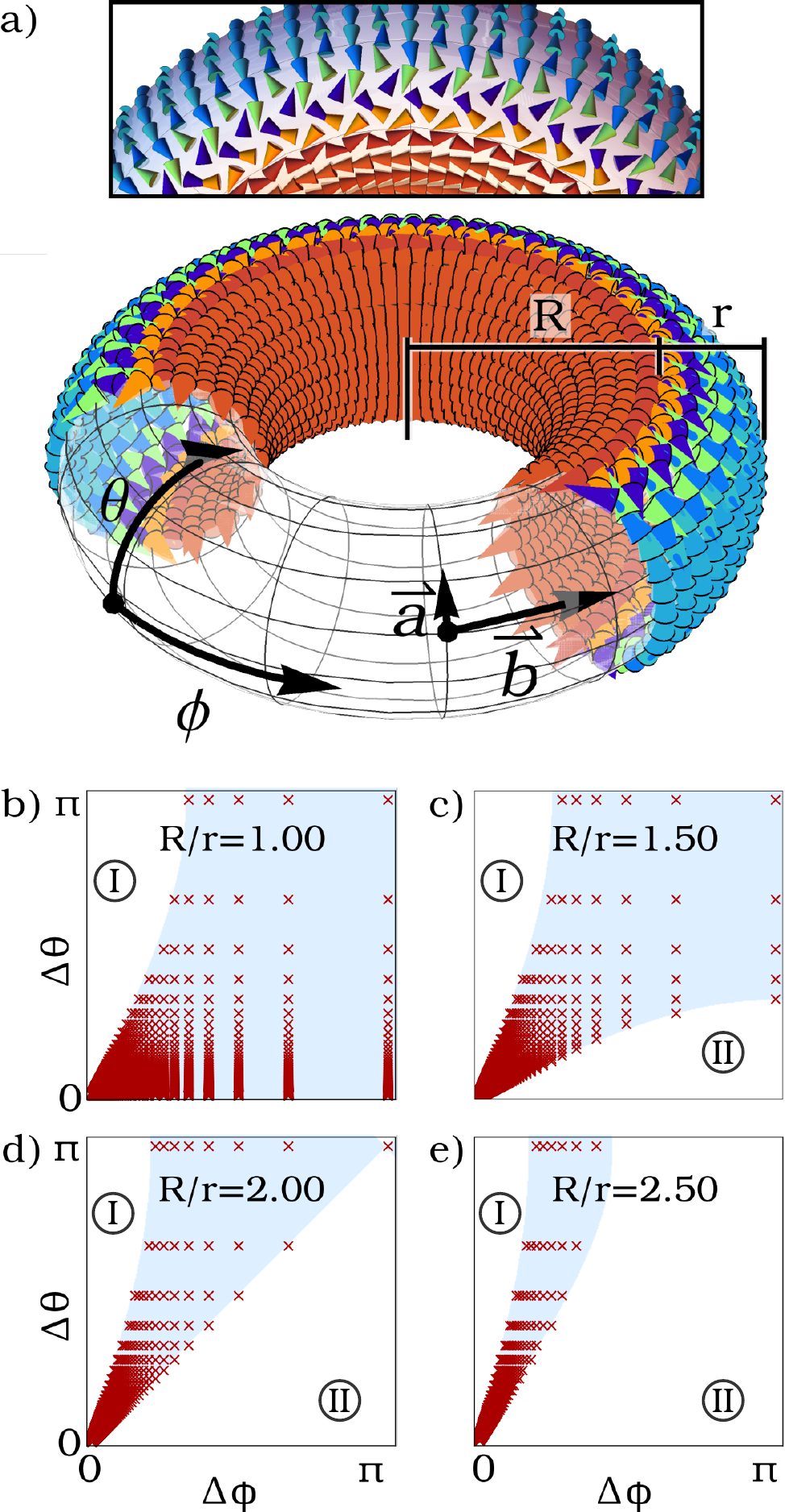}
\caption{\label{Fig:TorusSetup} (a) Schematic representation of the toroidal lattice geometry and parameters, together with the GS for a dipole lattice with $R/r=2.5$, $\Delta\theta=2\pi/30$, and $\Delta\phi=2\pi/80$. Dipole orientations are shown with colored cones. Coloring varies from red (dipoles aligned along $\phi$) to blue (dipoles aligned along $\theta$). The upper panel depicts a top-view to better showcase the dipole orientations. For reasons of illustration the lattice constants chosen for the schematic do not exactly correspond to the shown GS. (b-d) Cross sections through the parameter space for various values of $R/r$. Inside the blue colored regions the inequality defined in Eq. \ref{eq:torus_inequality} is fulfilled. Red crosses mark valid parameter combinations (i.e. parameters that fulfill the toroidal boundary conditions) for which we expect a DW in the GS. }
\end{figure}

In the previous section, we have shown how the GS configuration of a deformed dipole lattice can be predicted based on the ratio $\gamma=a/b$. 
To demonstrate that this behavior is of a more general character and can also be observed in compact curved or deformed dipole lattices, we now demonstrate the formation of domains and DWs for a square dipole lattice spanned on the surface of a torus. 
A schematic of the setup is provided in Fig. \ref{Fig:TorusSetup}(a) together with a corresponding GS dipole configuration. 
The torus surface can be described by the following parametric function
\begin{equation}\label{eq:1}
f_{t}(\theta,\phi):=\left(
\begin{array}{c}
\left(R+r\cos(\theta)\right)\cos(\phi) \\
\left(R+r\cos(\theta)\right)\sin(\phi) \\
r\sin(\theta)
\end{array}
\right),
\theta,\phi\in [0,2\pi]
\end{equation}
where $R$ and $r$ respectively describe the major and minor radius of the torus. 
Due to the periodic boundary conditions enforced by the toroidal geometry, the lattice constants $\Delta\theta$ and $\Delta\phi$ for a lattice with dimension $(N\times M)$ are given by 
\begin{equation}\label{eq:2}
\begin{array}{c}
\Delta\theta = 2\pi / (N+1) \\
\Delta\phi = 2\pi / (M+1)
\end{array}
\end{equation}
Similar to the previous example system, the euclidean distance of lattice points is constant along $\textbf{\^{e}}_{\theta}$, whereas the lattice point distance along $\textbf{\^{e}}_{\phi}$ depends on $\theta$.
Specifically, the euclidean distance $b$ between lattice points along the major torus radius direction increases with increasing distance from the torus center. 
The distances $a$ and $b$ can be analytically expressed as follows
\begin{equation}
\eqalign{
a^2 &= 2 r^2 \left[ 1-\cos(\Delta\theta) \right] \cr
b^2 (\theta) &= 2 (R + r \cos(\theta))^2 \left[ 1-\cos(\Delta\phi) \right] 
}
\end{equation}
The positions of the lattice points of a lattice with dimension ($N\times M$) are then given by $f_{t}(i\Delta\theta,k\Delta\phi)$ for $i\in[1,N]$ and $k\in[1,M]$.

We start by discussing the parameter values for which DWs and domains appear on the torus. 
In the limit of $a<b$ (i.e. $\gamma<1$), the dipoles will align in an AFE configuration parallel to $\textbf{\^{e}}_{\theta}$. 
In the limit of $a>b$ (i.e. $\gamma>1$), the dipoles will align in an AFE configuration parallel to $\textbf{\^{e}}_{\phi}$. 
The presence of both domains is expected when $\textrm{Min}[b(\theta)]<a<\textrm{Max}[b(\theta)]$. 
Since the maximum and minimum of $b$ are reached for $\theta=0$ and $\theta=\pi$ respectively, the condition for the presence of two domains can be expressed as
\begin{equation}\label{eq:torus_inequality}
\left(\frac{R}{r}-1\right)^2 < \frac{\left[1-\cos(\Delta\theta)\right]}{\left[1-\cos(\Delta\phi)\right]} < \left(\frac{R}{r}+1\right)^2
\end{equation}
It should be noted that fulfilling the above inequality is a necessary, but not a sufficient condition for the presence of two domains in the GS. 
In case the area on the torus where $\gamma>1$ (or $\gamma<1$) tends to zero (this will happen for $a\searrow \textrm{Min}[b(\theta)]$ or $a\nearrow \textrm{Max}[b(\theta)]$) the GS will feature only a single domain - even though the inequality Eq. \ref{eq:torus_inequality} is fulfilled. 

The inequality Eq. \ref{eq:torus_inequality} depends on $\Delta\theta$, $\Delta\phi$, and the ratio of the torus radii $R/r$. 
Figures \ref{Fig:TorusSetup}(b-e) show for $R/r\in[1,1.5,2,2.5]$ the combinations of $\Delta\theta$ and $\Delta\phi$ for which Eq. \ref{eq:torus_inequality} is fulfilled (blue areas). 
Values in the regime $R/r<1$ are not considered, since they correspond to configurations where the torus surface intersects itself. 
The subset of parameters that satisfy the toroidal boundary conditions are marked by red crosses in Fig. \ref{Fig:TorusSetup}(b-e).
An example for a ground state configuration for $R/r=2.5$, $\Delta\theta=2\pi/30$, and $\Delta\phi=2\pi/80$ (well within the region specified by Eq. \ref{eq:torus_inequality}) can be seen in Fig. \ref{Fig:TorusSetup}(a). 
Configurations outside of the blue regions in Figs. \ref{Fig:TorusSetup}(b-e) consist of a single domain of dipoles covering the complete torus and being aligned either along $\textbf{\^{e}}_{\theta}$ or $\textbf{\^{e}}_{\phi}$ [regions marked I or II in Figs. \ref{Fig:TorusSetup}(b-e) respectively]. 
For configurations within the blue region, the position of the DW can be controlled through the choice of $\Delta\theta$ and $\Delta\phi$. 
For parameters within the blue region but close to the border with region I, a narrow domain of dipoles aligned along $\textbf{\^{e}}_{\theta}$ will appear on the outer side of the torus. 
In the blue region and close to the border with region II the domain of dipoles aligned along $\textbf{\^{e}}_{\theta}$ will encompass almost the entire torus, whereas the domain of dipoles aligned along $\textbf{\^{e}}_{\phi}$ will only consist of a narrow band around the torus center.

The DW separating the two domains in the toroidal dipole lattice is reminiscent of the DW described in Sec. \ref{Sec:2D_lattice}: 
the DW separates regions with $\gamma=a/b>1$ from regions with $\gamma<1$. 
As one may expect, the orientation of the dipoles in the DW is also similar to the 2D case of Sec. \ref{Sec:2D_lattice}: 
the dipoles in the DW align in a `zig-zag' pattern similar to what we have observed in Fig. \ref{Fig:2dSetup}(b). 
Indeed, the order parameter $\chi$ (along a one-dimensional cut along $\textbf{\^{e}}_{\theta}$) can be well-described by a sigmoid function. 
However, in our toroidal dipole lattice, we notice a deviation from the behavior of the DW in the 2D lattice: 
The DW center is not exactly at the position where $\gamma=1$, but rather shifted w.r.t. it; for the example system shown in Fig. \ref{Fig:TorusSetup}(a), it is located at $\gamma\approx0.96$ - a difference of about $\Delta\theta/2$ on the surface.

\begin{figure}
\includegraphics[width=\columnwidth]{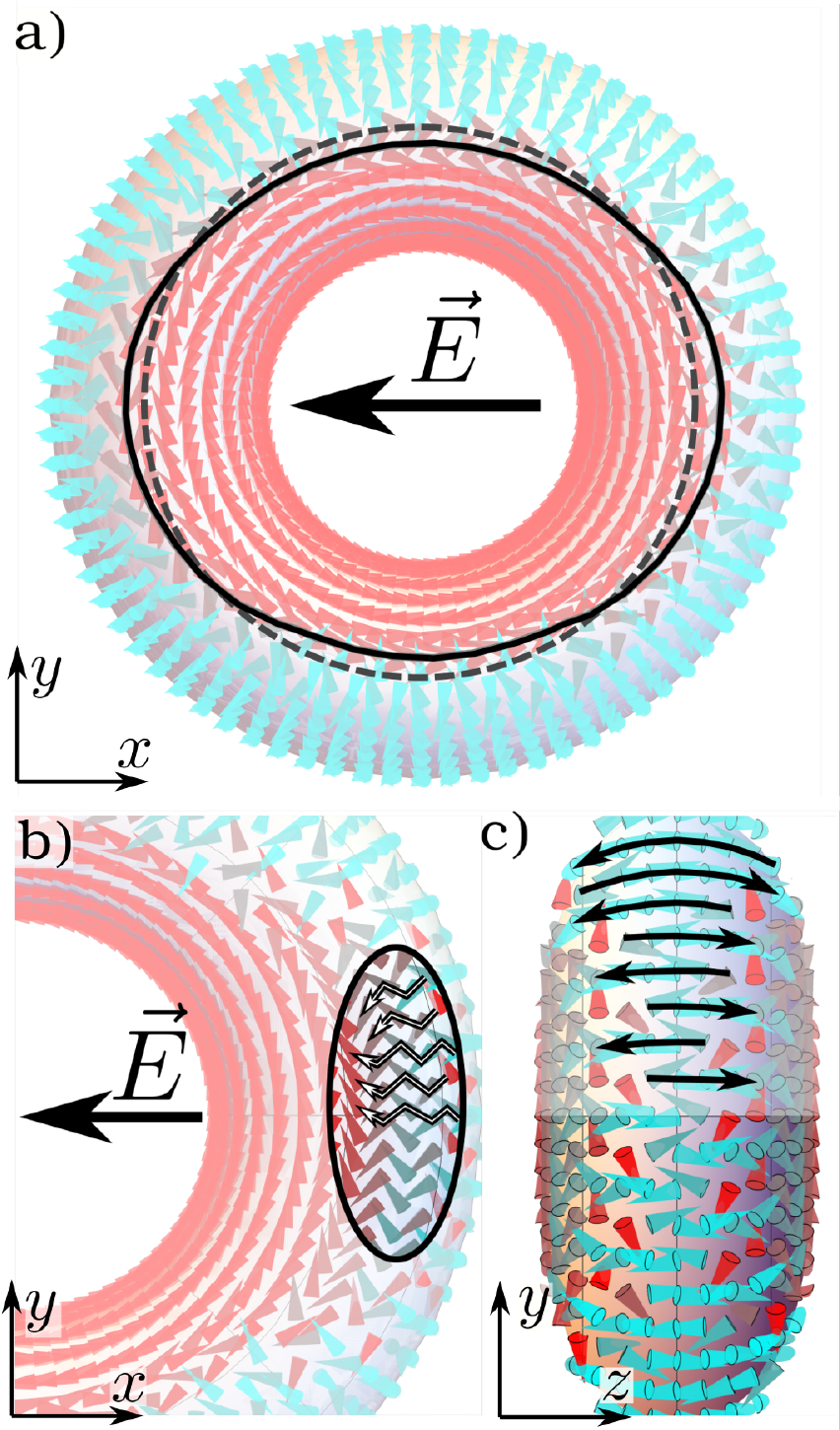}
\caption{\label{Fig:TorusField} Equilibrium configurations for the toroidal dipole lattice in the regime of (a) weak and (b-c) strong external electric fields. Coloring indicates orientation along $\textbf{\^{e}}_{\theta}$ (blue) and $\textbf{\^{e}}_{\phi}$ (red). (a) Weak electric fields cause an elliptic deformation of the DW-region. The solid (dashed) line indicate the position of the DW center for weak fields (no field). (b-c) Top and side view for a strong field. Dipoles in the highlighted region in (b) align along parallel zig-zag chains. These zig-zag chains will straighten into FE head-to-tail chains when the field strength is further increased. The side-view in (c) shows the FE domain on the outside of the torus. 
}
\end{figure}

In addition to the ground state properties, the corresponding response to an external electric field can also be observed in our toroidal dipole lattice. 
We would like to stress that the purpose of investigating the influence of electric fields is to highlight the difference to other (e.g. degeneracy induced) DWs in ferroelectrics and not an attempt at an in-depth study of the field response in itself. 
We consider therefore the configuration shown in Fig. \ref{Fig:TorusSetup}(a) and apply a field along the torus plane (x-direction). 
Simulation results for the case of such a field are shown in Fig. \ref{Fig:TorusField}. 
Just like for the flat lattice of Sec. \ref{Sec:2D_lattice}, a weak field causes a shift of the DW center towards the region where dipoles are aligned (anti-) parallel to the field. 
This is shown in Fig. \ref{Fig:TorusField}(a), where the position of the DW center is indicated by the solid line on the torus surface. 
To highlight the deformation, the position of the DW center without external field is indicated by the gray dotted line. 
In the regime of strong fields, the emergence of new FE domains around $\gamma=1$ can also be observed. 
These new domains emerge in those regions where the DW center has shifted the most from its original (field-free) position and expand from there when the field strength is increased. 
The formation of one of these FE domains is visualized in Fig. \ref{Fig:TorusField}(b). 
In the figure, the field strength does not quite suffice to force perfect FE order: 
rather than perfect head-to-tail chains, the dipoles in the highlighted region align in a zig-zag pattern along parallel chains (see black arrows in Fig. \ref{Fig:TorusField}(b)). 
Even larger field strengths are required for these zig-zag chains to straighten into a head-to-tail alignment. 
However, for the considered parameters any further increase of the field strength will result in the entire outer domain aligning parallel to the field.

\section{Domain-localized Excitations}\label{Sec:Excitations}
\begin{figure}
\includegraphics[width=\columnwidth]{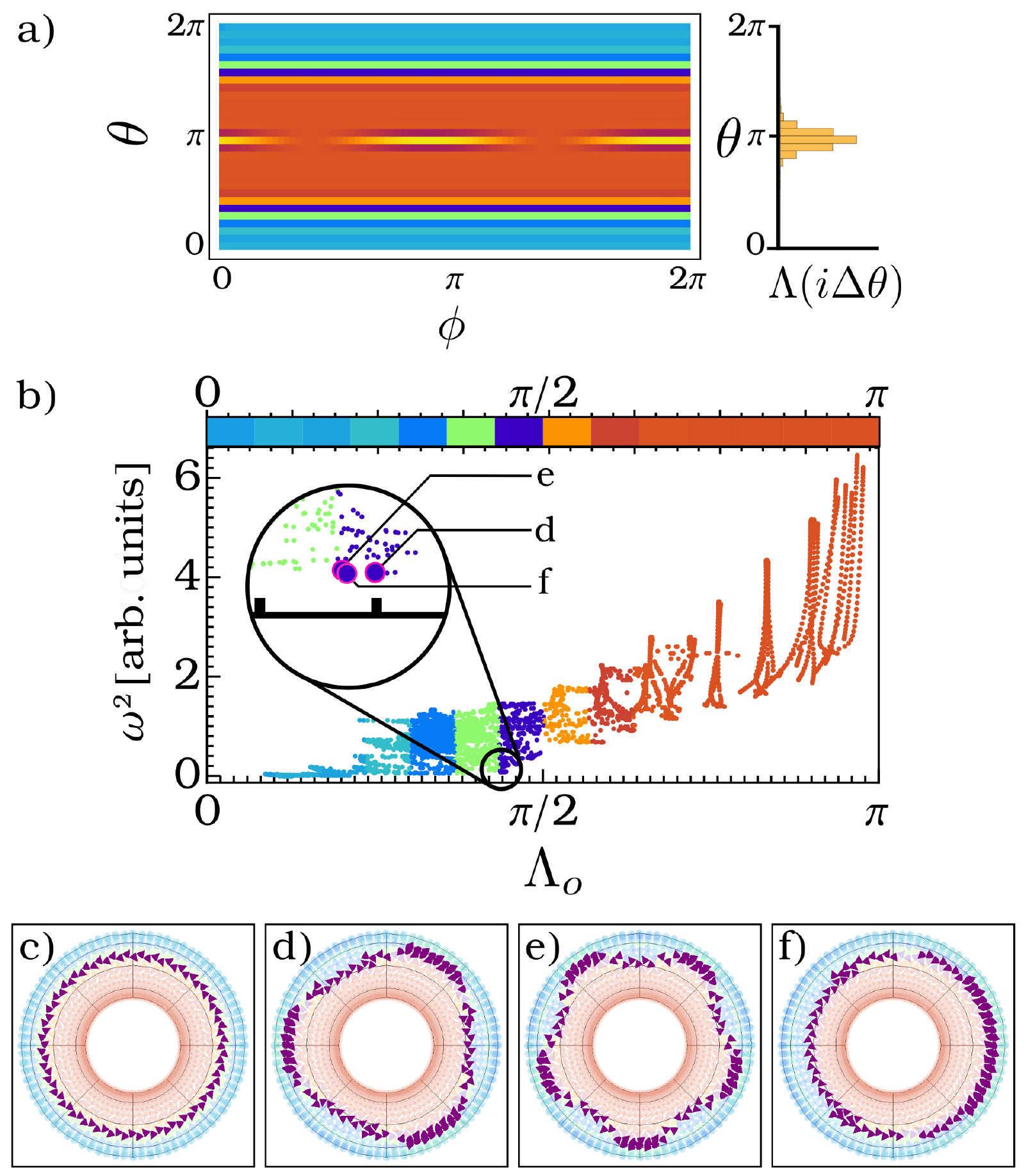}
\caption{\label{Fig:Eigenfrequencies} (a) Example visualization of the dipole orientations when an eigenmode is excited. 
Each position $(\theta, \phi)$ is colored according to the orientation of the closest dipole. 
The color-function is the same as in Fig. \ref{Fig:TorusSetup}, with blue and red colors respectively corresponding to dipoles oriented along $\textbf{\^{e}}_{\theta}$ and $\textbf{\^{e}}_{\phi}$. 
The dipoles that are significantly excited by the eigenmode, located in the inner domain at $\theta\approx\pi$, appear as yellow islands. 
The corresponding distribution $\Lambda(i\Delta\theta)$ on the right has been determined according to Eq. \ref{eq:Lambda}. (b) Squared eigenfrequencies $\omega^2$ as a function of $\Lambda_o$ (defined as the mean position of $\Lambda(i\Delta\theta)$). Coloring corresponds to the position of $\Lambda_o$ on the torus: Red (blue) colored points indicate that $\Lambda_o$ is located in the inner (outer) domain. (c) Top-view of the ground state of a toroidal dipole lattice with $r=0.4$, $\Delta\theta=2\pi/30$ and $\Delta\phi=2\pi/80$. (d-f) Top-views of eigenmodes corresponding to excitations of the domain-wall of the state shown in (c). The frequencies and localization of the eigenmodes shown in (d-f) are also highlighted in the zoom-in in (b). 
}
\end{figure}
We investigate in the following the dynamics of low energy excitations of the dipole orientations for the toroidal dipole lattice. 
We will demonstrate a tendency of these excitations to remain confined within the boundaries of a domain, i.e., these excitations will not spread or finally cover the entire torus. 
Low energy excitations are treated on basis of a harmonic approximation to the total potential energy using NN interactions. 
Specifically, the potential energy landscape around the ground state is approximated with a second order Taylor series. 
The first order terms of the series vanish by definition, since the ground state corresponds to a minimum of the potential. 
Consequently, only the second order terms will contribute to the equations of motion. 
The orientation of each dipole in euclidean space is described by local polar coordinates, i.e., a polar and azimuthal angle denoted $\mu$ and $\nu$, respectively. 
We will use the shorthand notation $\textbf{X}=\left(\mu_1,...,\mu_K,\nu_1,...,\nu_K\right)$ to describe the dipole configuration. 
Using this notation, the ground state will be referred to as $\textbf{X}_0=\left(\mu_1^0,...,\mu_K^0,\nu_1^0,...,\nu_K^0\right)$. 
In this case the linearized equations of motion for the polar and azimuthal angle of the $n$-th dipole are approximated as follows
\begin{equation}\label{eq:dyn_eom}
\eqalign{
	d^2\mu_n/dt^2 &= \sum_{i}^{K} \left[ \frac{\partial^2 V_n(\textbf{X})}{\partial\nu_i\partial\mu_n}\Big|_{\textbf{X}_0}\nu_i + \frac{\partial^2 V_n(\textbf{X})}{\partial\mu_i\partial\mu_n}\Big|_{\textbf{X}_0}\mu_i \right] \\
	d^2\nu_n/dt^2 &= \sum_{i}^{K} \left[ \frac{\partial^2 V_n(\textbf{X})}{\partial\nu_n\partial\mu_i}\Big|_{\textbf{X}_0}\mu_i + \frac{\partial^2 V_n(\textbf{X})}{\partial\nu_i\partial\nu_n}\Big|_{\textbf{X}_0}\nu_i \right]
}
\end{equation}
where $n$ is the index of the dipole, and $V_n$ is the corresponding potential energy as defined in Eq. \ref{eq:potentialEnergy}. 
For small amplitude motion (with angles $\alpha\leq5^{\circ}$) around the ground state the electromagnetic fields induced by the dipole motion can be neglected - as long as the nearest neighbor distances are much smaller than $c\pi/2\alpha\omega$, where $c$ is the speed of light and $\omega$ is the frequency of the dipole. 
This condition is derived from the ratio between the radiated and the `static' components of the electric field of an oscillating dipole \cite{purcell2013}. 
To put that into context: if we assume a NN distance on the nanometer scale, the emitted electromagnetic wave can be neglected as long as the dipole oscillation are frequencies much smaller than $\omega\ll10^{19}\, \mbox{s}^{-1}$. 
Assuming only low energy excitations, the solutions to Eq. \ref{eq:dyn_eom} take the form of spatiotemporal harmonic oscillations, allowing the problem to be written as an eigenvalue equation of the form 
\begin{equation}
\frac{d^2}{dt^2} \left(\begin{array}{c}
\vec{\mu} \\
\vec{\nu}
\end{array}\right)=
\omega^2 \left(\begin{array}{c}
\vec{\mu} \\
\vec{\nu}
\end{array}\right)= \mathcal{H} \left(\begin{array}{c}
\vec{\mu} \\
\vec{\nu}
\end{array}\right) ,
\end{equation}
where $\mathcal{H}$ is the Hessian matrix of second derivatives of $V_n(\textbf{X})$ evaluated at $\textbf{X}_0$, and $\vec{\mu}=\lbrace\delta\mu_1,...,\delta\mu_{N\times M}\rbrace^T, \vec{\nu}=\lbrace\delta\nu_1,...,\delta\nu_{N\times M}\rbrace^T$ are vectors containing the deviations of polar and azimuthal angles from the GS equilibrium configuration for all dipoles in the lattice. 
Eigenmodes of the system then correspond to the eigenvectors of the Hessian: 
for each eigenmode, the components of the eigenvector describe the oscillation amplitude for each lattice point on the torus. 
The corresponding eigenfrequencies are given by the square root of the corresponding eigenvalues of the Hessian.

\begin{figure}
\includegraphics[width=\columnwidth]{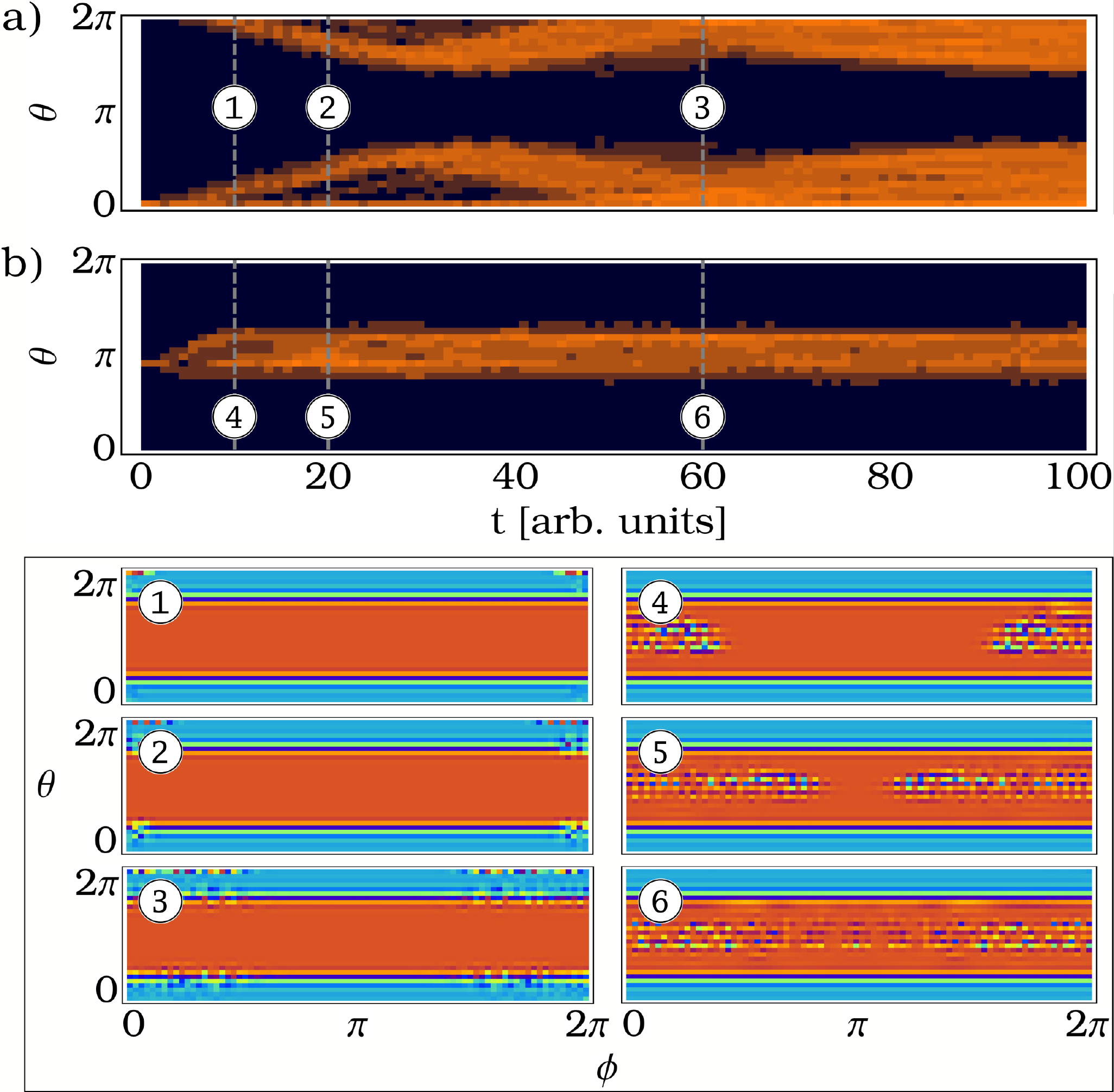}
\caption{\label{Fig:TimeEvolution} 
Time evolution an initial excitation at one lattice site in the outer domain (a) and in the inner domain (b). The coloring indicates the strength of the excitation. Dark colors indicate that the corresponding lattice sites are not perturbed and remain in their GS orientation (or very close to it). Bright colors indicate large amplitude deviations of the dipoles from the GS. 
The images marked 1-6 visualize the dipole orientations on the torus for different times $t\in[10,20,60]$. 
Images (1-3) and (4-6) respectively correspond to the excitations shown in (a) and (b). 
Just like in Fig. \ref{Fig:Eigenfrequencies}(a), each position $(\theta, \phi)$ is colored according to the orientation of the closest dipole. 
The color-function is the same as in Fig. \ref{Fig:TorusSetup}, with blue and red colors respectively corresponding to dipoles oriented along $\textbf{\^{e}}_{\theta}$ and $\textbf{\^{e}}_{\phi}$. 
}
\end{figure}

An example of an eigenmode that is confined entirely to the inner domain is shown in Fig. \ref{Fig:Eigenfrequencies}(a). 
In order to characterize the eigenmodes we define the alignment variation at each lattice point as the (absolute) angular rotation of the dipole from its ground state orientation 
$|\cos^{-1}\left[\textbf{d}(i\Delta\theta,k\Delta\phi)\cdot\textbf{d}_0(i\Delta\theta,k\Delta\phi)\right]|$. 
We can obtain a distribution $\Lambda(i\Delta\theta)$ of the eigenmode through summation of the alignment variation along the $\phi$-direction
\begin{equation}\label{eq:Lambda}
	\Lambda(i\Delta\theta)=\sum_{k} |\cos^{-1}\left[\textbf{d}(i\Delta\theta,k\Delta\phi)\cdot\textbf{d}_0(i\Delta\theta,k\Delta\phi)\right]|
\end{equation}
An example for such a distribution $\Lambda(i\Delta\theta)$ is shown on the right side of Fig. \ref{Fig:Eigenfrequencies}(a). 
The mean (designated $\Lambda_o$) and variance of the distribution $\Lambda(i\Delta\theta)$ provide a quantitative measure of the region to which an eigenmode is confined. 
Note that we define $\Lambda_o$ by the distance to the outermost part of the torus, such that $\Lambda_o=0$ ($\Lambda_o=\pi$) if the mean is located at the outermost (innermost) point of the torus. 
We analyzed the distributions $\Lambda(i\Delta\theta)$ for all eigenmodes and find that each eigenmode is confined to a specific region on the torus surface. 
This allows a classification of the eigenmodes into three types of excitations: 
excitations confined to the inner domain, the outer domain, or excitations of the DW. 
The $\Lambda_o$ of all eigenmodes are shown in Fig. \ref{Fig:Eigenfrequencies}(b), together with the corresponding eigenfrequencies. 
A discernible pattern is that $\Lambda_o$ increases with the eigenfrequency, i.e., low frequency modes have a $\Lambda_o$ located at the outer domain, whereas large frequency modes have a $\Lambda_o$ located at the inner domain. 
Eigenmodes with a $\Lambda_o$ located on the DWs can be found mainly for intermediate frequencies. 
The relation between $\omega$ and $\Lambda_o$ can be explained as follows: 
Compared to the outer domain, dipoles in the inner domain interact stronger with their neighbors (due to smaller NN distances). 
Consequently, dipoles in the outer domain require more energy to significantly change their orientation than dipoles in the outer domain. 
The result are low frequency oscillations in the outer, and large frequency oscillations in the inner domain. 
It is worth mentioning that excitations of the DW manifest mainly as sinusoidal deformations, resulting in star-shaped excitations on the torus, similar to those observed in e.g. Bose Einstein condensates \cite{kwon2021} or magnetic skyrmions \cite{rozsa2018}. 
Examples for these DW excitations are shown in Fig. \ref{Fig:Eigenfrequencies}(d-f).

The confinement of eigenmodes to certain regions of the torus surface effectively prevents arbitrary small energy excitations from exploring the torus beyond the boundary of the DW.
I.e. excitations started in the inner (outer) domain will never lead to a significant motion in the outer (inner) domain. 
This can be demonstrated by simulating the time evolution of perturbations of the GS. 
In practice, this is done by expressing an initial excitation as a linear combination of eigenmodes, and letting the eigenmodes oscillate with their respective eigenfrequencies. 
Specifically, using the notation $\textbf{E}_i, i\in\lbrace i|i\in\mathbb{Z},1\leq i\leq2K\rbrace$ to refer to the $2K$ eigenvectors, the time evolution of any initial small amplitude excitation $\delta\textbf{X}(t)=\textbf{X}_{ex}(t)-\textbf{X}_0$ can be expressed analytically as 
\begin{equation}
\delta\textbf{X}(t) = \sum_i^{2K} \cos(\omega_i t)\left[\delta\textbf{X}(0)\cdot \textbf{E}_i\right]\cdot \textbf{E}_i /E_i^2,  
\end{equation}
where $\omega_i$ is the eigenfrequency corresponding to the eigenvector $\textbf{E}_i$. 
In the following, we will consider two examples where the initial excitation $\delta\textbf{X}(0)$ is a unit vector with a single non-zero entry. 
This corresponds to an excitation a single dipole by rotating its polar angle by one radiant. 
We visualize the time evolution of excitations showing the distribution $\Lambda(i\Delta\theta,t)$ for discrete times $t$. 
Such a time evolution can be seen in Fig. \ref{Fig:TimeEvolution} for an initial excitation of a single dipole in the outer domain at $\theta=0$ (Fig. \ref{Fig:TimeEvolution}(a)) and an excitation of a single dipole in the inner domain at $\theta=39\pi/40$ 
(Fig. \ref{Fig:TimeEvolution}(b)). 
As one might expect, with increasing time the perturbation spreads through the system and an increasing number of dipoles begins to oscillate. 
However, contrary to what one might initially expect, the excitation does not cross over the DW into the other domain, but is reflected at the DW and thereby confined to the domain it originated from. 
The images (1-6) in the lower panel of Fig. \ref{Fig:TimeEvolution} visualize snapshots of the dipole configuration at different times $t\in[10,20,60]$. 
Images (1-3) correspond to the excitation shown in Fig. \ref{Fig:TimeEvolution}(a), whereas images (4-6) correspond to the excitation shown in Fig. \ref{Fig:TimeEvolution}(b). 
A comparison of images (1-3) and (4-6) shows that the excitation of the outer domain in (1-3) spreads slower than the excitation of the inner domain in (4-6). 
This can be understood by inspecting Fig. \ref{Fig:Eigenfrequencies}, which shows that eigenmodes that excite mainly the outer domain oscillate at lower frequencies than eigenmodes that excite predominantly the inner domain.

\section{Summary and Outlook}\label{Sec:Summary}
We have demonstrated that dipole lattices spanned on curved structures can exhibit behavior that drastically differs from that of dipole lattices in flat geometries. 
In particular, the (euclidean) distances between neighboring lattice points become dependent on the local geometry of the surface within which the lattice resides.
This can lead to the presence of anti-ferroelectric (AFE) domains in the ground state (GS) of the system. 
While the domain-walls (DW) separating these domains appear similar to other DWs found in ferroelectrics, their behavior can significantly differ. 
Most importantly, these geometry-induced DWs are `pinned' to positions where the ratio of the (euclidean) distances to the NNs is equal ($\gamma=1$). 
As a result of this `pinning' of the DW center to $\gamma=1$, these DWs will not simply shift when e.g. external electric fields are applied, bur rather show a more complex behavior: 
For low field amplitudes, the domain-wall position will be slightly displaced until a new static equilibrium is reached. 
For larger field amplitudes, a new FE domain will emerge between the two original domains. 
With increasing field amplitude, this FE domain will increasingly expand into the neighboring domains. 
Interestingly, this new FE domain appears before dipoles in the neighboring domains begin to significantly align with the field. 

For our prototype system of a toroidal dipole lattice, we further investigated the dynamics of low energy excitations of the GS. 
There, we demonstrated the tendency of DWs to inhibit the thermalization of the system by (effectively) preventing small energy excitations from exploring beyond the boundary of the DW. 
We expect this effect to be caused by the continuous change of the lattice constants: 
The dipoles towards the center of the torus are much closer to one another and therefore more `costly' to excite. 
Consequently, any excitation starting in the outer domain may be prevented by energetic considerations from exploring too far into any region where the dipoles are much closer and the interactions therefore much stronger. 
A similar effect of excitations being confined to certain regions has been described in certain soft matter systems and other inhomogeneous lattices \cite{zhang2021}. 

\ack{A. S. thanks A. Romero-Ros and M. Pyzh for helpful discussions. }

\bibliographystyle{vancouver}
\bibliography{txtest}

\end{document}